\documentclass[twocolumn,superscriptaddress,nobalancelastpage,pra]{revtex4}
\usepackage{amsmath}
\usepackage{amsthm}
\usepackage{amssymb}
\usepackage{amsfonts}
\usepackage{graphicx}
\usepackage{wasysym}
\usepackage{mathrsfs}
\usepackage{yfonts}
\usepackage{bbold}
\usepackage{verbatim}
\usepackage{subfigure}
\usepackage{color}

\newcommand{\ket}[1]{\left| #1 \right\rangle}
\newcommand{\bra}[1]{\left\langle #1 \right|}

\newcommand{\avgp}[1]{\langle #1 \rangle}

\newcommand{\opa}{\hat{a}}

\newcommand{\oEp}{\hat{E}^{(+)}}
\newcommand{\oEn}{\hat{E}^{(-)}}

\newcommand{\pher}{\Delta\phi}

\makeatletter
\newcommand*{\rom}[1]{\expandafter\@slowromancap\romannumeral #1@}
\makeatother

\makeatletter
\newcommand{\roml}[1]{\lowercase\expandafter{\romannumeral #1\relax}}
\makeatother

\DeclareMathAlphabet{\mathpzc}{OT1}{pzc}{m}{it}

\begin{document}

\title{Super-Sensitive Quantum Metrology with Separable States}
	
\author{Mayukh Lahiri}
\email{mlahiri@okstate.edu} \affiliation{Department of Physics, Oklahoma State University, Stillwater, OK 74078, USA.}
	\author{Manuel Erhard}
	\affiliation{Faculty  of  Physics, University of Vienna, Boltzmanngasse 5, Vienna A-1090, Austria.}
	\affiliation{Institute for Quantum Optics and Quantum Information, Austrian Academy of Sciences, Boltzmanngasse 3, Vienna A-1090, Austria.}
	
	\begin{abstract}
		We introduce a super-sensitive phase measurement technique that yields the Heisenberg limit without using either a squeezed state or a many-particle entangled state. Instead, we use a many-particle separable quantum state to probe the phase and we then retrieve the phase through single-particle interference. The particles that physically probe the phase are never detected. Our scheme involves no coincidence measurement or many-particle interference and yet exhibits phase super-resolution. We also analyze in detail how the loss of probing particles affects the measurement sensitivity and find that the loss results in the generation of many-particle entanglement and the reduction of measurement sensitivity. When the loss is maximum, the system produces a many-particle Greenberger–Horne–Zeilinger (GHZ) state, and the phase measurement becomes impossible due to very high phase uncertainty. In striking contrast to the super-sensitive phase measurement techniques that use entanglement involving two or more particles as a key resource, our method shows that having many-particle entanglement can be counterproductive in quantum metrology.
	\end{abstract}
	
	\maketitle
	
	Precise phase measurement is an important research topic in physics and has played a vital role in the famous Michelson-Morley experiment \cite{michelson1887relative} and in the first gravitational wave detection \cite{PhysRevLett.116.061102}. The topic is related to a fundamental question that is central to quantum metrology: How precisely can the phase be measured? Our understanding of quantum mechanics suggests that it is not possible to measure the phase ($\phi$) with infinite precision. When optical interferometry is used for phase measurement, the phase fluctuation or uncertainty ($\Delta\phi$) is affected by the photon number fluctuation ($\Delta n$). A seminal work of Dirac on the emission and absorption of radiation suggested that $\Delta\phi$ and $\Delta n$ should obey a Heisenberg-like uncertainty relation $\pher \Delta n \geq 1$  \cite{dirac1927quantum,Heitler-Q-rad}. Although the rigor of this relation has been debated (see, for example, \cite{Susskind-phase,barnett1989hermitian,MW}), there exists a common agreement that the precision of phase measurement depends on the (average) number ($n$) of photons or particles used in the measurement process. Sometimes $n$ is also understood as the number of interactions between the probe and the phase to be measured, i.e., the number of times the phase is sampled \cite{Giovannetti-nat-rev}. When the precision of phase measurement is limited by $\Delta \phi \geq n^{-1/2}$, the corresponding limit is called the standard quantum limit or shot-noise limit. Phase measurements performed with all classical interferometers are bound by the shot-noise limit. However, if quantum effects are used, more precise measurements can be performed. According to the current knowledge (see, for example, \cite{Caves-vs-Sh,Ou-supres,Ou-fund-ph-lim}), the minimum value of the phase uncertainty ($\pher$) is bounded by $1/n$ from below, i.e., $\pher\geq n^{-1}$. This limit is often called the Heisenberg limit. The goal of super-sensitive phase measurement is to achieve the Heisenberg limit or at least to beat the shot-noise limit.
	\par 
	A well-known method of enhancing phase sensitivity beyond the shot-noise limit involves the use of squeezed state of light \cite{Caves-sup-res-1,Kimble-supres}. This method has recently been applied to enhance sensitivity of the LIGO gravitational wave detector \cite{PhysRevLett.123.231107}. There is another class of methods that uses quantum entanglement involving two or more particles as a key resource to enhance phase sensitivity \cite{Yurke-sup-res-1,Yurke-sup-res-2,Dowling-PRA-1,lee2002quantum,kuzmich1998sub}. Entangled states involving two or more particles can also be used to achieve the phase super-resolution \cite{kuzmich1998sub,Fonseca-sup-res,Shih2001-sup-res,walther2004broglie,mitchell2004super}, which refers to a faster oscillation of the output signal of an interferometer with the change in phase. In such a case, it is a two- or many-particle interference pattern that exhibits phase-super resolution. 
	\par
	We introduce a fundamentally different method of super-sensitive and super-resolving phase measurement: we do not use any squeezed or many-particle entangled state to attain the Heisenberg limit; we instead use a many-particle quantum state in which \emph{no entanglement exists} between two or more particles. The presence of many-particle entanglement is disadvantageous for our scheme because it enhances the phase uncertainty. In fact, when the loss of probing particles is maximum, the quantum state becomes a Greenberger–Horne–Zeilinger (GHZ) state (maximally entangled) and the phase sensitivity is completely lost. Furthermore, in our scheme, the phase super-resolution is exhibited by a \emph{single-particle} interference pattern while we do not detect the particles that are used to probe the phase. 
	\par
	Our method is based on the concept of a recently introduced many-particle interferometer \cite{Lahiri-many-part-PI}. The theoretical analysis applies to both bosons and fermions.
	\begin{figure}[htbp]  \centering
		\includegraphics[width=0.98\linewidth]{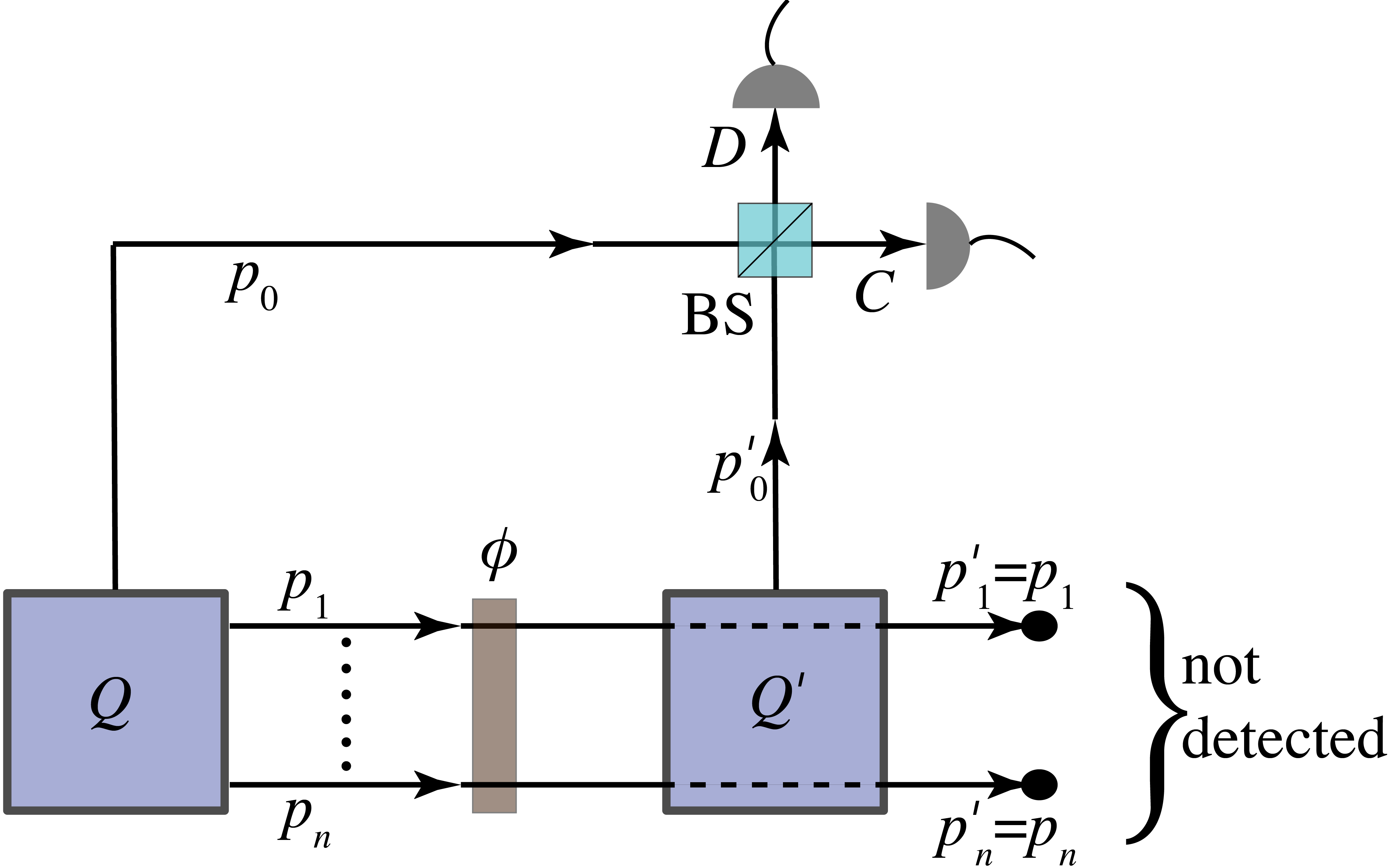}
		\qquad
		\caption{Super-sensitive and super-resolving phase measurement scheme. Two identical sources, $Q$ and $Q'$, can emit $n+1$ particles ($0,1,\dots, n$) into paths or modes $p_0,p_1, \dots, p_n$ and $p_0',p_1', \dots, p_n'$, respectively. Modes $p_1, \dots, p_n$ traverses a phase shifter that introduces phase $\phi$ into each mode. These modes are then sent through $Q'$ and made identical with modes $p_1', \dots, p_n'$. Particles $0,1,\dots, n$ are never detected. $Q$ and $Q'$ emit in such a way that the resulting quantum state is a superposition of the states that are generated by them individually. Modes $p_0$ and $p_0'$ are superposed by a beamsplitter, BS, and phase, $\phi$, is determined by performing measurements at output ports $C$ or $D$.} \label{fig:N-part-Pi-setup}
	\end{figure} 
	\par
	Our super-resolving and super-sensitive phase measurement scheme is illustrated in Fig. \ref{fig:N-part-Pi-setup}. We consider two \emph{identical} sources, $Q$ and $Q'$, each of which can emit $n+1$ particles (bosons or fermions). We label the particles by numerals $0,1,2, \dots, n$. Suppose that $Q$ can emit the particles in distinct paths or modes $p_0,p_1, \dots, p_n$. Therefore, if $Q$ emits individually, the corresponding quantum state is represented by 
	\begin{align}\label{state-Q}
	\ket{X}=\ket{p_0}\ket{p_1}\dots \ket{p_n}=\prod_{j=0}^n
	\opa^{\dag}(p_j)\ket{\text{vac}},
	\end{align}
	where $\opa^{\dag}(p_j)$ is the creation operator corresponding to particle $j$ in path (mode) $p_j$ and $\ket{\text{vac}}$ represents the vacuum state. The annihilation operator $\opa_{p_j}$ can be bosonic or fermionic. It obeys the commutation relation $\opa(p_j)\opa^{\dag}(p_k)\pm \opa^{\dag}(p_k)\opa(p_j)=\delta_{jk}$, where the minus and plus signs are for bosons and fermions, respectively and $\delta_{jk}$ represents the Kronecker delta. For fermions, the following formula holds in addition: $\{\opa(p_j)\}^s=\{\opa^{\dag}(p_j)\}^s=0$ for $s>1$, where $s$ is a positive integer. Likewise, $Q'$ can emit the particles in paths (modes) $p_0',p_1', \dots, p_n'$ and the corresponding individual quantum state is given by
	\begin{align}\label{state-Qp}
	\ket{X'}=\ket{p'_0}\ket{p'_1}\dots \ket{p'_n}=\prod_{j=0}^n
	\opa^{\dag}(p_j')\ket{\text{vac}}.
	\end{align} 
	\par
	The phase ($\phi$) that we wish to measure is introduced into each of the paths (modes) $p_1, \dots, p_n$ by a phase shifter \footnote{If the particles are photons, a phase shifter can be a glass plate of appropriate thickness.} that is placed between $Q$ and $Q'$ (Fig. \ref{fig:N-part-Pi-setup}). For bosons, these modes do not need to be distinct; however, for fermions, these modes must be distinct. Paths (modes) $p_1, \dots, p_n$ are then sent through $Q'$ and are made identical with paths (modes) $p_1', \dots, p_n'$, respectively \footnote{Making paths identical in this way was suggested by Z. Y. Ou for a signle-photon and was experimentally implemented in \cite{zou1991induced}.}. Such an arrangement can be called path identity \cite{krenn2017entanglement,Lahiri-many-part-PI} and be represented analytically by the relationship \cite{zou1991induced,lahiri2015theory,lahiri2017twin}
	\begin{align}\label{PI-no-loss}
	\opa^{\dag}(p_l')=e^{-i\phi}\opa^{\dag}(p_l), \quad l=1,2,\dots, n.
	\end{align} 
	\begin{figure*}[htbp]  \centering
		\includegraphics[width=1\linewidth]{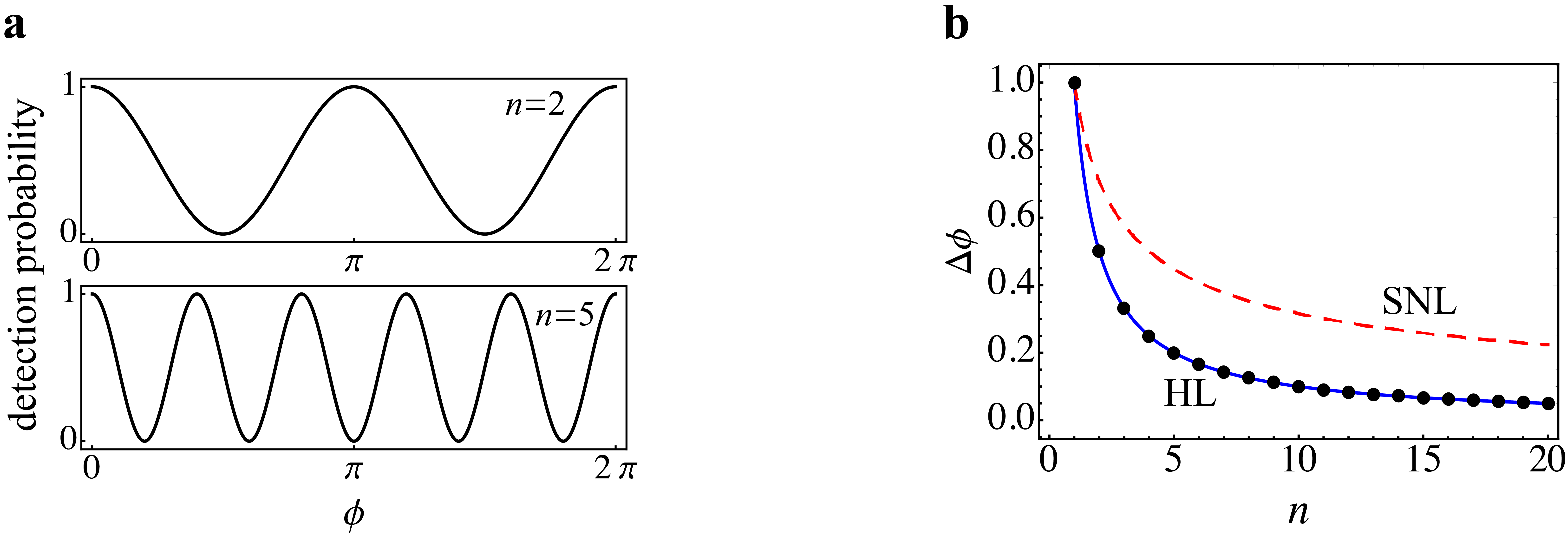}
		\qquad
		\caption{Simulated results demonstrating super-resolution and super-sensitivity. \textbf{a}, Super-resolution: the probability of single-particle detection at output $D$ [Eq. (\ref{sing-part-pattern:b})] is plotted against phase $\phi$ for different number of probing particles ($n=2$ and $n=5$). The detection probability oscillates between its maximum and minimum values more frequently for a larger value of $n$. \textbf{b}, Super-sensitivity: phase uncertainty $\pher$ is plotted against the number of probing particles $n$ (filled circles). The Heisenberg limit (HL) of $1/n$ (solid line) and shot-noise limit (SNL) of $1/\sqrt{n}$ (dashed line) are shown for comparison.} \label{fig:results}
	\end{figure*}
	\par
	The sources emit in such a way that the resulting state is a superposition of $\ket{X}$ and $\ket{X'}$. Combining Eqs. (\ref{state-Q}), (\ref{state-Qp}), and (\ref{PI-no-loss}), we now find that the quantum state generated by the system is
	\begin{align}\label{state-final}
	\ket{\psi}=\frac{1}{\sqrt{2}} \big(\ket{p_0}+e^{i(\xi-n\phi)} \ket{p_0'}\big)
	\prod_{l=1}^n \ket{p_l},
	\end{align}
	where $\xi$ is an arbitrary constant phase. It is evident from the quantum state, $\ket{\psi}$, that there is \emph{no entanglement} between two or more particles. We \emph{never} detect particles $1,2,\dots,n$ that are used to probe the phase. Equation (\ref{state-final}) shows that phase $\phi$ appears in the superposition of $\ket{p_0}$ and $\ket{p_0'}$ although particle ``$0$'' is not used to probe the phase. 
	\par
	We combine paths (modes) $p_0$ and $p_0'$ by a beamsplitter (BS) and detect both outputs $C$ and $D$ (Fig. \ref{fig:N-part-Pi-setup}). We denote the number operator at outputs $C$ and $D$ by $\hat{n}_C$ and $\hat{n}_D$, respectively. The \emph{difference number operator} is defined by $\hat{n}_\text{diff}=\hat{n}_D-\hat{n}_C$. The phase uncertainty is then given by the standard formula (see, for example, \cite{Yurke-sup-res-1,kuzmich1998sub})
	\begin{align}\label{phase-uncertainty-def}
	\pher=\Delta n_\text{diff}\left(\left|\frac{\partial}{\partial\phi}\avgp{\hat{n}_\text{diff}}\right|\right)^{-1},
	\end{align}
	where $\Delta n_\text{diff}=\sqrt{\avgp{\hat{n}_\text{diff}^2}-\avgp{\hat{n}_\text{diff}}^2}$ and the angular brackets represent the quantum mechanical average. 
	\par
	The quantum field at outputs $C$ and $D$ can be represented by \footnote{These formulas correspond to a photonic beamsplitter. Similar formulas apply to beamsplitters for fermions and other kinds of bosons.}
	\begin{subequations}\label{E-CD}
		\begin{align}
		\oEp_C &=\frac{1}{\sqrt{2}}\left[\opa(p_0)+ie^{i\gamma}\opa(p_0') \right], \label{E-CD:a} \\
		\oEp_D &=\frac{1}{\sqrt{2}}\left[i\opa(p_0)+e^{i\gamma}\opa(p_0') \right], \label{E-CD:b} 
		\end{align}
	\end{subequations}
	where $\gamma$ is the phase difference between paths $p_0$ and $p_0'$; we have chosen the beamsplitter to be lossless and symmetric for simplicity. It follows from Eqs. (\ref{state-final}) and (\ref{E-CD}) that the single-particle detection probability at outputs $C$ and $D$ are given by
	\begin{subequations}\label{sing-part-pattern}
		\begin{align}
		\avgp{\hat{n}_C} & =\bra{\psi}\oEn_C \oEp_C\ket{\psi}=\frac{1}{2} \left[ 1-\cos\left( n\phi-\zeta_0 \right) \right], \label{sing-part-pattern:a} \\
		\avgp{\hat{n}_D} & =\bra{\psi}\oEn_D \oEp_D\ket{\psi}= \frac{1}{2} \left[ 1+\cos\left( n\phi-\zeta_0 \right) \right],
		\label{sing-part-pattern:b}
		\end{align}
	\end{subequations}
	where $\oEn_C=\{ \oEp_C\}^{\dag}$, $\oEn_D=\{ \oEp_D\}^{\dag}$, and $\zeta_0=\xi+\gamma-\pi/2$. Clearly, single-particle interference patterns with \emph{unit visibility} are generated at outputs $C$ and $D$ when the phase is varied. Making the modes $p_1, \dots, p_n$ identical with the modes $p_1', \dots, p_n'$ ensures that there is no which-way information and thus the interference occurs. 
	\par
	The presence of $n\phi$ inside the cosines [Eq. (\ref{sing-part-pattern})] shows that the detection probabilities oscillate between their maximum and minimum values more frequently for a larger value of $n$. In Fig. \ref{fig:results}(a), we plot the single-particle detection probability at output $C$ against the phase $\phi$ for $n=2$ and $n=5$. It is evident that our phase measurement scheme exhibits phase super-resolution in single-particle interference without using two- or many-particle entanglement. This is a striking result because the phase super-resolution is commonly observed in many-particle interference experiments performed with entangled states involving two or more particles. \cite{kuzmich1998sub,Fonseca-sup-res,Shih2001-sup-res,walther2004broglie,mitchell2004super}. 
	\par
	We now proceed to determine the phase uncertainty ($\pher$). The difference number operator is given by 
	\begin{align}\label{diff-n-op}
	\hat{n}_\text{diff}=\hat{n}_D-\hat{n}_C=\oEn_D \oEp_D-\oEn_C \oEp_C.
	\end{align}
	From Eqs. (\ref{state-final}), (\ref{E-CD}), (\ref{sing-part-pattern}), and (\ref{diff-n-op}), we find for both bosons and fermions that \footnote{The calculations for Eq. (\ref{diff-n-avg-sd:a}) are identical for bosons and fermions. For Eq. (\ref{diff-n-avg-sd:b}), the only difference comes in the explicit form of the operator $\hat{n}_\text{diff}^2$: For bosons, $\hat{n}_\text{diff}^2=\hat{n}_{p_0}+\hat{n}_{p_0'}+2\hat{n}_{p_0}\hat{n}_{p_0'}+\{\opa^{\dag}(p_0)\}^2\{\opa(p_0')\}^2e^{2i(\gamma-\pi/2)}+\{\opa^{\dag}(p_0')\}^2\{\opa(p_0)\}^2e^{-2i(\gamma-\pi/2)}$, whereas for fermions, $\hat{n}_\text{diff}^2=\hat{n}_{p_0}+\hat{n}_{p_0'}-2\hat{n}_{p_0}\hat{n}_{p_0'}$; here $\hat{n}_{p_0}=\opa^{\dag}(p_0)\opa(p_0)$ and $\hat{n}_{p_0'}=\opa^{\dag}(p_0')\opa(p_0')$. However, $\avgp{\hat{n}_\text{diff}^2}=1$ for both cases and thus the equations hold for both bosons and fermions.}
	\begin{subequations}\label{diff-n-avg-sd}
		\begin{align}
		&\avgp{\hat{n}_\text{diff}} =\avgp{\hat{n}_D}-\avgp{\hat{n}_C}= \cos\left(n\phi-\zeta_0 \right), \label{diff-n-avg-sd:a} \\
		&(\Delta n_\text{diff})^2 =\avgp{\hat{n}_\text{diff}^2}-\avgp{\hat{n}_\text{diff}}^2= \sin^2\left(n\phi-\zeta_0\right).
		\label{diff-n-avg-sd:b}
		\end{align}
	\end{subequations}
	Using Eqs. (\ref{phase-uncertainty-def}) and (\ref{diff-n-avg-sd}), we readily find that the phase uncertainty \textemdash ~for both the bosonic and the fermionic cases \textemdash ~is given by 
	\begin{align}\label{phase-uncertainty-res}
	\pher=\frac{1}{n},
	\end{align}
	for all values of $\phi$. Equation (\ref{phase-uncertainty-res}) confirms that our scheme allows one to achieve the Heisenberg-limit of phase super-sensitivity. 
	\par
	In Fig. \ref{fig:results}(b), we plot $\pher$ against $n$ (black dots) to illustrate that the phase uncertainty obtained in our scheme attains the Heisenberg limit (solid line). In the same figure, we also show the shot-noise limit (dashed line) for comparison. 
	\par
	We have thus far assumed that there is no loss in the system. We now consider the loss of the probing particles and show why the presence of entanglement between two or more-particles is counterproductive for our phase measurement scheme. In our system, the loss can be effectively represented  by the action of an attenuator placed in the paths (modes) of the probing particles between the two sources $Q$ and $Q'$ \cite{Lahiri-many-part-PI}. The mathematical representation of such an attenuator is equivalent to that of a beamsplitter with one unused input port \cite{zou1991induced}. Let us denote the amplitude transmission coefficient of the attenuator by $T_j$ for mode $p_j$ ($j=1,2,\dots n$), where $0\leq T_j\leq 1$ can be assumed to be real without any loss of generality. Using the quantum mechanical treatment of a beamsplitter (\cite{MW}, Sec. 12.12), we now replace Eq. (\ref{PI-no-loss}) by
	\begin{align}\label{PI-loss}
	\opa^{\dag}(p_l')=e^{-i\phi}\left[T_l\opa^{\dag}(p_l)+\sqrt{1-T_l^2}\opa_0^{\dag}(p_l)\right],
	\end{align} 
	where $l=1,2,\dots, n$ and $\opa_0^{\dag}(p_l)$ corresponds to the vacuum field at the unused port of the beamsplitter that represents the attenuator. Note that  $1-T_l^2$ is the probability of particle $l$ getting lost and $\opa_0^{\dag}(p_l)\ket{\text{vac}}$ can be interpreted as the state representing a lost particle.
	\par
	Combining Eqs. (\ref{state-Q}), (\ref{state-Qp}), and (\ref{PI-loss}), we find that the quantum state generated by a lossy system can be expressed as
	\begin{align}\label{state-final-loss}
	&\ket{\psi_\text{loss}}=\frac{1}{\sqrt{2}}\ket{p_0}\prod_{l=1}^n \ket{p_l} \nonumber \\ &+\frac{1}{\sqrt{2}}e^{i(\xi-n\phi)} \ket{p_0'}
	\prod_{l=1}^n \left(T_l\opa^{\dag}(p_l)+R_l\opa_0^{\dag}(p_l) \right) \ket{\text{vac}},
	\end{align}
	where $R_l=\sqrt{1-T_l^2}$ and $\bra{\text{vac}} \opa_0(p_l) \opa_0^{\dag}(p_l)\ket{\text{vac}}=1$. Depending on the value of $n$, the state $\ket{\psi_\text{loss}}$ is a two- or many-particle entangled state. It now follows from Eqs. (\ref{E-CD}), (\ref{diff-n-op}), and (\ref{state-final-loss}) that for both bosons and fermions \footnote{Once again, $\avgp{\hat{n}_\text{diff}^2}=1$ for both bosons and fermions; see [36] in this context.},
	\begin{subequations}\label{diff-n-avg-sd-loss}
		\begin{align}
		&\avgp{\hat{n}_\text{diff}} =\cos\left(n\phi-\zeta_0 \right)\prod_{l=1}^nT_l, \label{diff-n-avg-sd-loss:a} \\
		&(\Delta n_\text{diff})^2 =1-\cos^2\left(n\phi-\zeta_0\right)\big(\prod_{l=1}^nT_l\big)^2,
		\label{diff-n-avg-sd-loss:b}
		\end{align}
	\end{subequations}
	where $\zeta_0$ is defined below Eq. (\ref{sing-part-pattern}). It is evident from Eq. (\ref{diff-n-avg-sd-loss:a}) that the visibility of the single-particle interference pattern is now given by 
	\begin{align}\label{vis-loss}
	\mathcal{V}=\prod_{l=1}^nT_l,
	\end{align}
	where $0\leq T_l \leq 1$. Clearly, the loss of probing particles can be quantified by the visibility.
	\begin{figure}[htbp]  \centering
		\includegraphics[width=0.98\linewidth]{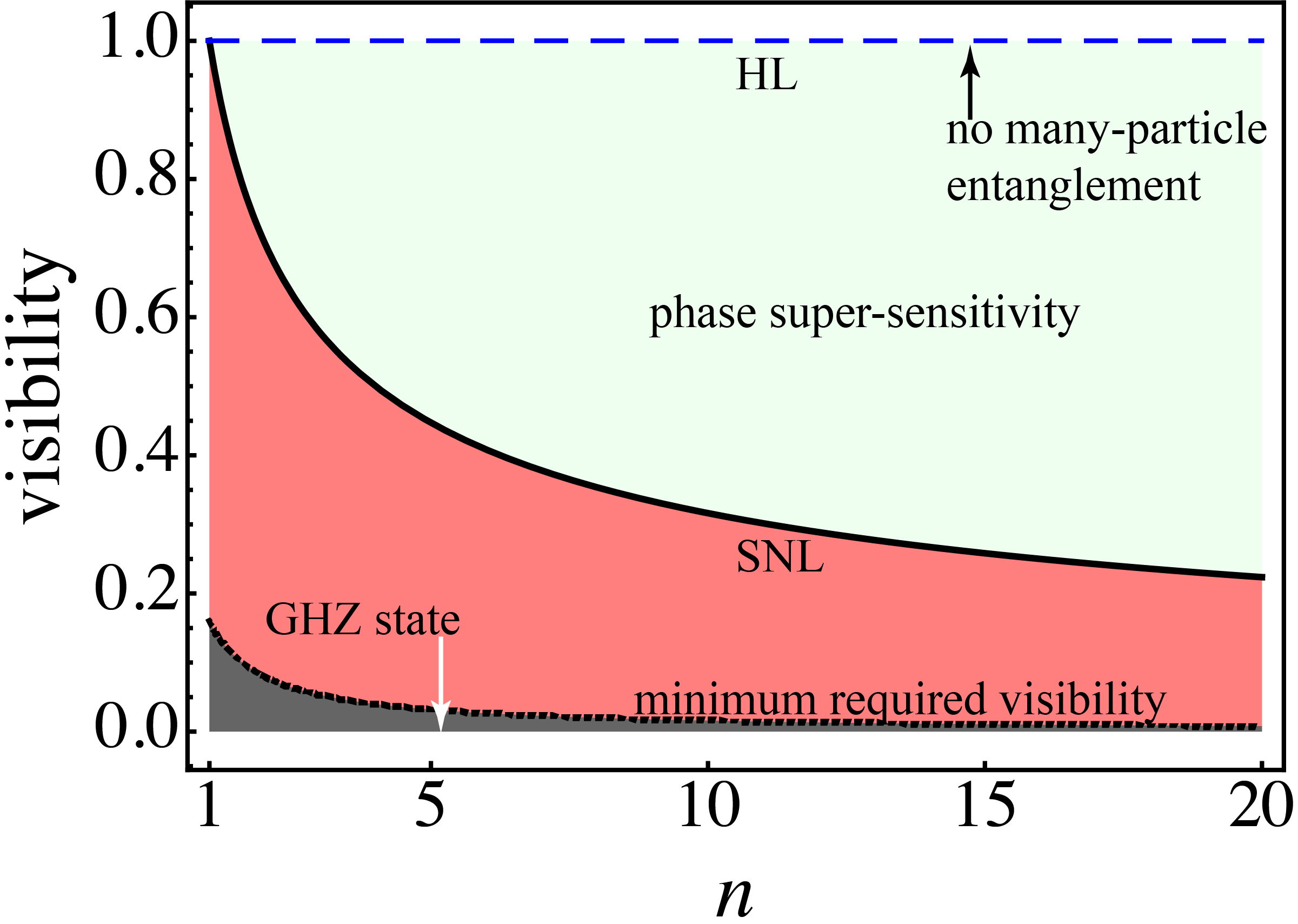}
		\qquad
		\caption{Phase sensitivity, loss, and entanglement. The Heisenberg limit (HL) is attained when there is no loss; in this case, there is no many-particle entanglement and the visibility is $1$ (dashed line). For a lossy system, the visibility is less than $1$, the phase sensitivity reduces, and the quantum state is entangled. Visibility greater than $1/\sqrt{n}$ (above the solid line) ensures better sensitivity than shot-noise limit (SNL). The visibility must be greater than $(2n\pi)^{-1}$ (dotted line) for a feasible phase measurement. For $100\%$ loss, the visibility is $0$ and the quantum state is a GHZ state; in this case, no phase measurement is possible.} \label{fig:min-vis-req}
	\end{figure}   
	\par
	From Eqs. (\ref{phase-uncertainty-def}) and (\ref{diff-n-avg-sd-loss}), we find that for a lossy system, the phase uncertainty is given by (for both bosons and fermions)
	\begin{align}\label{phase-uncertainty-res-loss}
	\pher_{\text{loss}}=\frac{1}{n}\left\{ \frac{1-\cos^2(n\phi-\zeta_0)\left(\prod_{l=1}^nT_l\right)^2}{\sin^2(n\phi-\zeta_0)\left(\prod_{l=1}^nT_l\right)^2} \right\}^{\frac{1}{2}}.
	\end{align}
	We note that in contrast to the case of lossless system, the phase uncertainty now depends on the value of $\phi$. The minimum value of the phase uncertainty is obtained when $n\phi-\zeta_0=\pi/2+m\pi$, where $m=\pm 1, \pm 2, \dots$. From Eqs. (\ref{vis-loss}) and (\ref{phase-uncertainty-res-loss}), we find that the minimum phase uncertainty for a lossy system is
	\begin{align}\label{phase-uncertainty-min-loss}
	\pher_{\text{loss}}\big|_{\text{min}}=\frac{1}{n\prod_{l=1}^nT_l}=\frac{1}{n\mathcal{V}}.
	\end{align}
	Therefore, when $\mathcal{V}>1/\sqrt{n}$ and $n>1$, a lossy system implementing our scheme will exhibit better phase sensitivity than the shot-noise limit exhibited by an ideal (lossless) classical system. Let us also note that the phase is usually defined modulo $2\pi$ and thus the maximum value of $\pher$ must be bounded by $2\pi$ from above. It means that when the visibility is less than $(2n\pi)^{-1}$, no realistic phase measurement is possible. Figure \ref{fig:min-vis-req} illustrates these results. The Heisenberg limit is achieved when the visibility is unity (i.e., no loss) as suggested by Eqs. (\ref{phase-uncertainty-res}) and (\ref{phase-uncertainty-res-loss}).
	\par
	For achieving the Heisenberg limit, it is absolutely essential that there is no entanglement between two or more particles [Eq. (\ref{state-final})]. The loss of probing particles introduces distinguishability and consequently there is many-particle entanglement [Eq. (\ref{state-final-loss})]. In such a case, the visibility of the interference pattern reduces [Eq. (\ref{vis-loss})] and the phase uncertainty ($\pher$) increases [Eq. (\ref{phase-uncertainty-min-loss})]. In the extreme case of modes $p_1,\dots, p_n$ being fully distinguishable from modes $p_1',\dots, p_n'$ (i.e., 100\% loss), the quantum state becomes a GHZ state (maximally entangled) \footnote{We ignore the trivial case of $n=1$ because it is not possible to achieve either super-resolution or super-sensitivity in this case.}. Consequently, the visibility becomes zero and the phase measurement can no longer be performed (Fig. \ref{fig:min-vis-req}). Clearly, many-particle entanglement reduces the phase-sensitivity of our measurement scheme: the presence of entanglement is counterproductive. This fact marks a striking difference between our scheme and the super-sensitive phase measurement schemes that use entanglement involving two or more particles (e.g., NOON state) as a key resource.
	\par
	We conclude by stating that we have introduced a unique method in quantum metrology. A distinct feature of our method is that the probing particles are not detected. Our scheme also does not involve any coincidence measurement or many-particle interference. The detection of single particles is only required and the phase super-resolution is observed in a single-particle interference pattern. Importantly, our scheme allows one to achieve the Heisenberg limit without using two- or many-particle entanglement as a resource. In fact, we have shown that the presence of many-particle entanglement is disadvantageous because it reduces the sensitivity of measurement [38]. The Heisenberg limit is attained only when the \emph{probing particles} are in a separable state.
	\par
	An intriguing question that arises in the context of our results concerns the role played by entanglement in important areas of quantum technology. In quantum computing, it is debated whether many-particle entanglement is a key resource for the computational speed-up \cite{gottesman1998heisenberg,Jozsa-noent-comp,Nest-noent-comp}. In quantum communication, measurement-device-independent quantum key distribution can be performed without using entangled photon pairs \cite{Lo-noent-QKD}. Now, our results show that it is possible to achieve the Heisenberg limit in quantum metrology without using a many-particle entangled state; a path is thus opened up for more questions about entanglement from a resource theoretical perspective. We, therefore, hope that our results will stimulate further discussion on this topic and help us develop a deeper understanding of quantum physics as a whole.
	\par
	\emph{Acknowledgements}.\textemdash ~The authors thank Professor Anton Zeilinger for enriching discussions. The research was partly supported by the Austrian Science Fund (FWF) with SFB F40 (FOQUS) and W 1210-N25 (CoQuS), and by the University of Vienna via the project QUESS. M.L. also acknowledges support from the College of Arts and Science and the Office of Vice President for Research, Oklahoma State University.
	
	\bibliography{references-supres.bib}
	
\end{document}